\documentclass[9pt,twocolumn,a4paper]{extarticle}
\usepackage{graphicx}
\usepackage{comment}
\usepackage{appendix}
\usepackage[font=footnotesize,format=plain,labelfont=bf,up]{caption}
\usepackage{subcaption}
\usepackage{sidecap}
\usepackage{cancel}
\usepackage{authblk}
\usepackage{slashed}
\usepackage{bm}
\usepackage{amstext}
\usepackage{amsmath}			
\usepackage{amssymb}
\usepackage{wasysym}
\usepackage{setspace}

\usepackage[utf8]{inputenc}
\usepackage[francais,british]{babel}     
\usepackage{tabularx}				
\usepackage{psfrag}
\usepackage{siunitx}
\usepackage[explicit,toctitles,pagestyles]{titlesec}
\usepackage{pifont}
\usepackage{csquotes}
\usepackage{varwidth}
\usepackage{lipsum}
\usepackage[numbers]{natbib}
\usepackage{color}
\usepackage{xcolor}
\usepackage{titletoc}
\usepackage{tikz}
\usepackage{tikz-3dplot} 
\tdplotsetmaincoords{60}{110}
\pgfmathsetmacro{\rvec}{.8}
\pgfmathsetmacro{\thetavec}{30}
\pgfmathsetmacro{\phivec}{60}
\usetikzlibrary{decorations.pathmorphing}
\usetikzlibrary{shapes}

\definecolor{link}{rgb}{0,0,.75}
\definecolor{cite}{rgb}{.75,0,0}
\definecolor{url}{rgb}{0,.75,0}
\usepackage[pdftex,colorlinks=false,pdfstartview=FitV,linkbordercolor=link,citebordercolor=cite,urlbordercolor=url,hyperindex=true,hyperfigures=false]{hyperref}

\let\oldsqrt\sqrt
\def\sqrt{\mathpalette\DHLhksqrt}
\def\DHLhksqrt#1#2{
  \setbox0=\hbox{$#1\oldsqrt{#2\,}$}\dimen0=\ht0
  \advance\dimen0-0.2\ht0
  \setbox2=\hbox{\vrule height\ht0 depth -\dimen0}
  {\box0\lower0.4pt\box2}}
\def\XXint#1#2#3{{\setbox0=\hbox{$#1{#2#3}{\int}$}
\vcenter{\hbox{$#2#3$}}\kern-.5\wd0}}

\makeatletter

\newcommand{\Rmnum}[1]{\expandafter\@slowromancap\romannumeral #1@}
\makeatother

\begin{document}
\title{Spontaneous light emission by atomic Hydrogen: Fermi's golden rule without cheating}
\author{V. Debierre, T. Durt, A. Nicolet and F. Zolla}
\affil{Aix Marseille Université, CNRS, École Centrale de Marseille, Institut Fresnel\\
UMR 7249, 13013 Marseille, France.}
\date{\today}
\twocolumn[
\maketitle
\vglue -1.8truecm
\vspace{20pt}
\begin{abstract}
Focusing on the $2\mathrm{p}-1\mathrm{s}$ transition in atomic Hydrogen, we investigate through first order perturbation theory the time evolution of the survival probability of an electron initially taken to be in the excited ($2\mathrm{p}$) state. We examine both the results yielded by the standard dipole approximation for the coupling between the atom and the electromagnetic field -for which we propose a cutoff-independent regularisation- and those yielded by the exact coupling function. In both cases, Fermi's golden rule is shown to be an excellent approximation for the system at hand: we found its maximal deviation from the exact behaviour of the system to be of order $\SI{d-8}{}/\SI{d-7}{}$. Our treatment also yields a rigorous prescription for the choice of the optimal cutoff frequency in the dipole approximation. With our cutoff, the predictions of the dipole approximation are almost indistinguishable at all times from the exact dynamics of the system.\\
\end{abstract}

\vspace{20pt}
]

\section{Introduction} \label{sec:Intro}

The description of spontaneous light emission by atomic electrons is a staple of textbooks in quantum physics \cite{Cohen2,Messiah2,LeBellac} and quantum optics \cite{FoxMulder}. Oftentimes, several approximations are made in the treatment. The first one is the dipole approximation \cite{Messiah2,FoxMulder} which consists \cite{Shirokov} in considering that the decaying electron does not emit light from its own position but rather from the position of the nucleus to which it is bound. This approximation is usually justified by claiming that the electromagnetic wavelengths which are relevant to the problem are much larger than the uncertainty on the electron's position. It is thus argued that the electromagnetic field does not ``see'' the details of matter configuration at the atomic scale, and hence that the precise location of the point of light emission is irrelevant. The second approximation is often used to explain why the dipole approximation can be made. It consists in noticing that, for large enough times, the only field modes which effectively contribute to spontaneous emission are the ones which are resonant with the atomic transition frequency. This is known as Fermi's golden rule. These resonant modes have a wavelength which is indeed much larger than the relevant atomic dimensions, and their interaction with the atom is very well described in the framework of the dipole approximation. Hence to some extent the dipole approximation is justified by Fermi's golden rule. But it is a very general result \cite{MisraSudarshan} of quantum physics that Fermi's linear decay cannot be valid at very short times. Indeed very short times obey what is called the Zeno dynamics, where the decay is always quadratic. One should therefore be skeptical of the validity of the dipole approximation at very short times, for which it may not be justified by the golden rule. Elsewhere \cite{EdouardArXiv} two of us have investigated the Hydrogen $2\mathrm{p}-1\mathrm{s}$ transition numerically, and developed a numerical method which enabled us to reproduce the dynamics of the system in the Zeno and Fermi regimes, but also at longer times in the Wigner-Weisskopf regime.\\

In the following we investigate the validity of the dipole and Fermi approximations in the case of the $2\mathrm{p}-1\mathrm{s}$ transition in atomic Hydrogen. In sect.~\ref{sec:AtomQED} we recall the main tools needed for the description of spontaneous emission. In sect.~\ref{sec:Heart}, the central section of this manuscript, we show how Fermi's golden rule emerges from rigorous first order time-dependent perturbation theory, using two new independent arguments. First, in the dipole approximation, which is discussed at length in sect.~\ref{subsec:DipoleDsc}, we regularise the divergences (sect.~\ref{subsec:DipoleCalc}) obtained in the expression for the survival probability of the excited state. The regularisation procedure is cutoff-independent. Then (sect.~\ref{subsec:MultipoleCalc}) we go beyond the dipole approximation and rigorously derive the short-time dynamics of the sytem.

\section{The decay of a two-level atom} \label{sec:AtomQED}

\subsection{Position of the problem} \label{subsec:Notations}

We consider a two-level atom, where the ground state $\mid\!\mathrm{g}\rangle$ has angular frequency $\omega_{\mathrm{g}}$ and the excited state state $\mid\!\mathrm{e}\rangle$ has angular frequency $\omega_{\mathrm{e}}$, interacting with the electromagnetic field in the rotating wave approximation. We call $e$ the (positive) elementary electric charge, $m_e$ is the electron mass, and $\hat{\mathbf{x}}$ is the position operator and $\hat{\mathbf{p}}$ the linear momentum operator for the electron. The atom is considered to be in free space. The Hamiltonian $\hat{H}=\hat{H}_A+\hat{H}_R+\hat{H}_I$ is a sum of three terms: the atom Hamiltonian $\hat{H}_A$, the electromagnetic field Hamiltonian $\hat{H}_R$, and the interaction Hamiltonian $\hat{H}_I$. In the Schrödinger picture these read \cite{CohenQED1}
\begin{subequations} \label{eq:Hamilton}
                \begin{align}
                  \hat{H}_A&=\hbar\omega_{\mathrm{g}}\mid\!\mathrm{g}\rangle\langle \mathrm{g}\!\mid + \hbar\omega_{\mathrm{e}}\mid\!\mathrm{e}\rangle\langle \mathrm{e}\!\mid&\\
                  \hat{H}_R&=\sum_{\lambda=\pm}\int\tilde{\mathrm{d}k}\,\hbar c\left|\left|\mathbf{k}\right|\right|\hat{a}_{\left(\lambda\right)}^\dagger\left(\mathbf{k}\right)\hat{a}_{\left(\lambda\right)}\left(\mathbf{k}\right)\\
                  \hat{H}_I&=\frac{e}{m_e}\hat{\mathbf{A}}\left(\hat{\mathbf{x}},t=0\right)\cdot\hat{\mathbf{p}}
                \end{align}
              \end{subequations}
where $\lambda$ labels the polarisation of the electromagnetic field. Introduce the polarisation vectors $\bm{\epsilon}_{\left(\lambda=\pm1\right)}\left(\mathbf{k}\right)$, which are any two (possibly complex) mutually orthogonal unit vectors taken in the plane orthogonal to the wave vector $\mathbf{k}$. The vector potential $\hat{\mathbf{A}}\left(\hat{\mathbf{x}},t\right)$ is expanded (in the Coulomb gauge) over plane waves as
\begin{equation} \label{eq:MaxwellSolHat}
\mathbf{\hat{A}}\left(\mathbf{x},t\right)=\sqrt{\frac{\hbar}{\epsilon_0c}}\sum_{\lambda=\pm}\int\tilde{\mathrm{d}k}\left[\hat{a}_{\left(\lambda\right)}\left(\mathbf{k}\right)\bm{\epsilon}_{\left(\lambda\right)}\left(\mathbf{k}\right)\mathrm{e}^{-\mathrm{i}\left(c\left|\left|\mathbf{k}\right|\right|t-\mathbf{k}\cdot\mathbf{x}\right)}+\mathrm{h.c.}\right].
\end{equation}
Here
\begin{equation} \label{eq:InDaTilde}
\tilde{\mathrm{d}k}\equiv\frac{\mathrm{d}^4k}{\left(2\pi\right)^4}2\pi\,\delta\left(k_0^2-\mathbf{k}^2\right)\theta\left(k_0\right),
\end{equation}
is the usual volume element on the light cone \cite{Weinberg1,ItzyksonZuber} (where $\theta$ stands for the Heaviside distribution). Finally, the commutation relation between the photon ladder operators is given by
\begin{equation} \label{eq:aaDagger}
\left[\hat{a}_{\left(\varkappa\right)}\left(\mathbf{k}\right),\hat{a}_{\left(\lambda\right)}^\dagger\left(\mathbf{q}\right)\right]=2\left|\left|\mathbf{k}\right|\right|\left(2\pi\right)^3\delta\left(\mathbf{k}-\mathbf{q}\right)\delta_{\varkappa\lambda}.
\end{equation}
The state of the system reads
\begin{multline}\label{eq:Ansatz}
          \mid\!\psi\left(t\right)\rangle=c_{\mathrm{e}}\left(t\right)\mathrm{e}^{-\mathrm{i}\omega_{\mathrm{e}}t}\mid\!\mathrm{e},0\rangle\\
          +\sum_{\lambda=\pm}\int\tilde{\mathrm{d}k}\,c_{\mathrm{g},\lambda}\left(\mathbf{k},t\right)\mathrm{e}^{-\mathrm{i}\left(\omega_{\mathrm{g}}+ c\left|\left|\mathbf{k}\right|\right|\right) t}\mid\!\mathrm{g},1_{\lambda,\mathbf{k}}\rangle
        \end{multline}
where $\mid\!\mathrm{e},0\rangle$ means that the atom is in the excited state and the field contains no photons and $\mid\!\mathrm{g},1_{\lambda,\mathbf{k}}\rangle$ means that the atom is in the ground state and the field contains a photon of wave vector $\mathbf{k}$ and polarisation $\lambda$.

\subsection{Importance of the interaction matrix element} \label{subsec:StartingHere}

In the absence of any interaction (that is, if $\hat{H}_I$ were zero), the coefficients $c_{\mathrm{e}}$ and $c_{\mathrm{g},\lambda}\left(\mathbf{k},\cdot\right)$ would not evolve and each term of the superposition (\ref{eq:Ansatz}) would oscillate at its own eigenfrequency. As such, the nontrivial features of the problem are encompassed by the matrix elements of the interaction Hamiltonian in the Hilbert (sub)space spanned by $\mid\!\mathrm{e},0\rangle$ and $\mid\!\mathrm{g},1_{\lambda,\mathbf{k}}\rangle$. In the dipole approximation, one approximates the imaginary exponential in (\ref{eq:MaxwellSolHat}) as $\mathrm{e}^{\mathrm{i}\mathbf{k}\cdot\mathbf{x}}\simeq1$ (remember that it is the vector potential at $t=0$ which appears in $\hat{H}_I$, reflecting the fact that none of the Hamiltonians in (\ref{eq:Hamilton}) depend explicitly on time). Then one is left to compute the relatively simple matrix elements of $\hat{\mathbf{p}}$. The result is well-known for the $2\mathrm{p}-1\mathrm{s}$ Hydrogen transition. Writing, for this transition, $\mid\!\mathrm{g}\rangle\equiv\mid\!1\mathrm{s}\rangle$ and $\mid\!\mathrm{g}\rangle\equiv\mid\!2\mathrm{p}\,m_2\rangle$, with $m_2$ the magnetic quantum number of the $2\mathrm{p}$ sublevel considered, we have the following expressions for the electronic wave functions of the $1\mathrm{s}$ and $2\mathrm{p}\,m_2$ sublevels:
\begin{subequations} \label{eq:HLevels}
\begin{align}
\psi_{1\mathrm{s}}\left(\mathbf{x}\right)&=\frac{\exp\left(-\frac{\left|\left|\mathbf{x}\right|\right|}{a_0}\right)}{\sqrt{\pi a_0^3}},\\
\psi_{2\mathrm{p}\,m_2}\left(\mathbf{x}\right)&=\frac{\exp\left(-\frac{\left|\left|\mathbf{x}\right|\right|}{2a_0}\right)}{8\sqrt{\pi a_0^3}}\frac{\sqrt{2}}{a_0}\mathbf{x}\cdot\bm{\xi}_{m_{2}}.
\end{align}
\end{subequations}
with $a_0$ the Bohr radius. The vectors $\bm{\xi}_{m_{2}}$ give the preferred directionality of the wave function of the $2\mathrm{p}$ substates, the angular dependence of which is given by the usual spherical harmonics \cite{Cohen1}. They are given by
\begin{subequations} \label{eq:Xi}
\begin{align}
\bm{\xi}_0&=\mathbf{e}_z,\\
\bm{\xi}_{\pm1}&=\mp\frac{\mathbf{e}_x\pm\mathrm{i}\mathbf{e}_y}{\sqrt{2}}.
\end{align}
\end{subequations}
In the dipole approximation, the interaction matrix element then is \cite{FacchiPhD}
\begin{subequations} \label{eq:MatrixDiMulti}
\begin{equation} \label{eq:MatrixDipole}
\langle1\mathrm{s},1_{\lambda,\mathbf{k}}\mid\!\hat{H}_I\!\mid\!2\mathrm{p}\,m_2,0\rangle=-\mathrm{i}\sqrt{\frac{\hbar}{\epsilon_0 c}}\frac{\hbar e}{m_e\,a_0}\frac{2^{\frac{9}{2}}}{3^4}\bm{\epsilon}_{\left(\lambda\right)}^*\left(\mathbf{k}\right)\cdot\bm{\xi}_{m_2}.
\end{equation}
The dipole approximation is further discussed in sect.~\ref{subsec:DipoleDsc}. For the same transition, if one keeps the full exponential in (\ref{eq:MaxwellSolHat}), one obtains the exact matrix element \cite{FacchiPhD}
\begin{equation} \label{eq:MatrixMultipole}
\langle1\mathrm{s},1_{\lambda,\mathbf{k}}\mid\!\hat{H}_I\!\mid\!2\mathrm{p}\,m_2,0\rangle=-\mathrm{i}\sqrt{\frac{\hbar}{\epsilon_0 c}}\frac{\hbar e}{m_e\,a_0}\frac{2^{\frac{9}{2}}}{3^4}\frac{\bm{\epsilon}_{\left(\lambda\right)}^*\left(\mathbf{k}\right)\cdot\bm{\xi}_{m_2}}{\left[1+\left(\frac{2}{3}a_0\left|\left|\mathbf{k}\right|\right|\right)^2\right]^2}.
\end{equation}
\end{subequations}
Since we are interested in spontaneous emission, we set $c_{\mathrm{e}}\left(t=0\right)=1$ and $\forall\lambda\in\left\{1,2\right\}\forall\mathbf{k}\in\mathbb{R}^3c_{\mathrm{g},\lambda}\left(\mathbf{k},t=0\right)=0$. We want to compute quantities such as
\begin{subequations} \label{eq:OntoCardSine}
\begin{equation} \label{eq:DecaytoSomeMode}
P_{\mathrm{emiss.}\rightarrow\lambda,\mathbf{k}}\left(t\right)=\left|c_{\mathrm{g},\lambda}\left(\mathbf{k},t\right)\right|^2=\left|\langle\mathrm{g},1_{\lambda,\mathbf{k}}\!\mid\!\hat{U}\left(t\right)\!\mid\!\mathrm{e},0\rangle\right|^2
\end{equation}
where $\hat{U}\left(t\right)=\exp\left[\left(-\mathrm{i}/\hbar\right)\hat{H}t\right]$ is the evolution operator for the system. It is well-known \cite{Cohen2,Englert} that a time-dependent perturbative treatment of such a problem yields, to first order in time,
\begin{equation} \label{eq:WildCardSineAppears}
P_{\mathrm{emiss.}\rightarrow\lambda,\mathbf{k}}\left(t\right)=\frac{t^2}{\hbar^2}\left|\langle\mathrm{g},1_{\lambda,\mathbf{k}}\!\mid\!\hat{H}_I\!\mid\!\mathrm{e},0\rangle\right|^2\sin\!\mathrm{c}^2\left[\left(\omega_0-c\left|\left|\mathbf{k}\right|\right|\right)\frac{t}{2}\right]
\end{equation}
\end{subequations}
where we introduced the notation $\omega_0\equiv\omega_{\mathrm{e}}-\omega_{\mathrm{g}}$. From (\ref{eq:Ansatz}) we deduce that the probability that, at time $t$, the electron is still in the excited state, which we shall call the survival probability, is given by
\begin{equation} \label{eq:CardSineSurv}
\begin{aligned} [b]
P_{\mathrm{surv}}\left(t\right)\equiv\left|c_{\mathrm{e}}\left(t\right)\right|^2&=1-\sum_{\lambda=\pm}\int\tilde{\mathrm{d}k}\left|c_{\mathrm{g},\lambda}\left(\mathbf{k},t\right)\right|^2\\
&=1-\frac{t^2}{\hbar^2}\sum_{\lambda=\pm}\int\tilde{\mathrm{d}k}\left|\langle\mathrm{g},1_{\lambda,\mathbf{k}}\!\mid\!\hat{H}_I\!\mid\!\mathrm{e},0\rangle\right|^2\\
&\hspace{75pt}\sin\!\mathrm{c}^2\left[\left(\omega_0-c\left|\left|\mathbf{k}\right|\right|\right)\frac{t}{2}\right].
\end{aligned}
\end{equation}
The integral in (\ref{eq:CardSineSurv}) will be central for the rest of the article. Before we investigate it in further detail, let us remind the reader of the usual \cite{Cohen2,Englert,LeBellac} treatment of the problem. It consists in using the distributional limit
\begin{equation} \label{eq:CardinalSineDirac}
\lim_{a\to+\infty}\frac{\sin^2\left(ax\right)}{a^2\,x}=\pi\delta\left(x\right)
\end{equation}
of the square cardinal sine, to conclude that the only single-photon states available by spontaneous emission have a frequency equal to the atomic transition frequency $\omega_0$. Indeed the use of (\ref{eq:CardinalSineDirac}) in (\ref{eq:WildCardSineAppears}) yields Fermi's golden rule
\begin{equation} \label{eq:OnlyResonant}
P_{\mathrm{emiss.}\rightarrow\lambda,\mathbf{k}}\left(t\right)\underset{t\rightarrow+\infty}{\sim}2\pi\frac{t}{\hbar^2}\left|\langle\mathrm{g},1_{\lambda,\mathbf{k}}\!\mid\!\hat{H}_I\!\mid\!\mathrm{e},0\rangle\right|^2\delta\left(\omega_0-c\left|\left|\mathbf{k}\right|\right|\right).
\end{equation}
Notice that making use of (\ref{eq:CardinalSineDirac}) in (\ref{eq:CardSineSurv}) means taking the limit $t\rightarrow+\infty$, a strange manipulation in the framework of time-dependent perturbation theory around $t=0$. One then has to enter a subtle discussion of time regimes to conclude that Fermi's golden rule is valid for times large enough to guarantee that one can approximate the square cardinal sine by its limit (\ref{eq:CardinalSineDirac}), but small enough to ensure that first-order perturbation theory still applies. This is discussed in more detail in \cite{LeBellac} for instance.\\

Our goal is not merely to point out the well-known fact that the golden rule is an approximation, but to acquire knowledge on what it is an approximation of. This is the question we investigate in the following sect.~\ref{sec:Heart}, where we focus on the case of the $2\mathrm{p}-1\mathrm{s}$ transition in atomic Hydrogen.

\section{Rigorous first order perturbation theory and Fermi's golden rule} \label{sec:Heart}

We shall investigate corrections to Fermi's golden rule from first order time-dependent perturbation theory. In sect.~\ref{subsec:DipoleDsc} we discuss the validity of the ubiquitous dipole approximation for the atom-field coupling. In sect.~\ref{subsec:DipoleCalc} we propose a cutoff-independent regularisation procedure of the divergences which arise when the dipole approximation is made. We then go on to investigate the dynamics yielded by the exact coupling in sect.~\ref{subsec:MultipoleCalc}.

\subsection{The dipole approximation: discussion} \label{subsec:DipoleDsc}

It has been noticed that in the framework of the dipole approximation, it is necessary to introduce a cutoff over electromagnetic field frequencies. In our treatment, this can be seen as follows: write the survival probability of the electron in the excited state at time $t$ as given by (\ref{eq:MatrixDipole}) and (\ref{eq:CardSineSurv}). Performing the integration over the angles, we get, using (\ref{eq:InDaTilde}), the following expression:
\begin{equation} \label{eq:NoChanceSurvival}
P_{\mathrm{surv}}\left(t\right)=1-\frac{2^{10}}{3^{9}\pi}c^2\,\alpha^3\,t^2\int_0^{+\infty}\mathrm{d}\omega\,\omega\sin\!\mathrm{c}^2\left(\left(\omega_0-\omega\right)\frac{t}{2}\right)
\end{equation}
where $\alpha=e^2/\left(4\pi\epsilon_0\hbar c\right)$ is the fine structure constant. The integral on the right-hand side of (\ref{eq:NoChanceSurvival}) diverges at all times. In this situation, taking the limit $t\rightarrow+\infty$ under the integral to obtain Fermi's golden rule is nonrigorous to say the least. Accordingly, one introduces a cutoff frequency corresponding to an upper bound on the validity domain of the dipole approximation for the atom-field coupling. One then considers that electromagnetic field modes with frequency larger than this cutoff frequency are uncoupled to the atom. Several objections can be raised with regard to this procedure. We discuss them along with their more or less satisfying rebuttals:
\begin{itemize}
\item{In order to introduce a cutoff, one must assume that high-frequency electromagnetic modes do not interact with the atom. This is justified by a quick look at the exact coupling. The matrix elements are proportional to $\langle1\mathrm{s}\!\mid\hat{\mathbf{p}}\exp\left(\mathrm{i}\mathbf{k}\cdot\hat{\mathbf{x}}\right)\mid\!2\mathrm{p}\,m_2\rangle$. When the wave number $\left|\left|\mathbf{k}\right|\right|$ becomes higher than the inverse of the Bohr radius $a_0$, such matrix elements are de facto negligibly small because the oscillating exponential $\exp\left(\mathrm{i}\mathbf{k}\cdot\hat{\mathbf{x}}\right)$ averages out during the integration, as seen from (\ref{eq:HLevels}).}
\item{The approximation yields cutoff-dependent results. Since the qualitative argument above only provides an order of magnitude (the ratio $c/a_0$) for the cutoff frequency $\omega_{\mathrm{C}}$, this is especially problematic. Indeed, one can check that the truncated integral
\begin{multline} \label{eq:Trunk}
\int_0^{\omega_{\mathrm{C}}}\mathrm{d}\omega\,\omega\sin\!\mathrm{c}^2\left(\left(\omega_0-\omega\right)\frac{t}{2}\right)\\=\frac{2}{t^2}\left[\log\left(-1+\frac{\omega_{\mathrm{C}}}{\omega_0}\right)-\left[\mathrm{Ci}\left(\left(\omega_{\mathrm{C}}-\omega_0\right)t\right)-\mathrm{Ci}\left(\omega_0 t\right)\right]\right.\\\left.+\frac{\omega_0}{\omega_{\mathrm{C}}-\omega_0}\left(-1+\cos\left(\left(\omega_{\mathrm{C}}-\omega_0\right)t\right)\right)+\left(-1+\cos\left(\omega_0 t\right)\right)\right.\\\left.+t\left(\mathrm{Si}\left(\left(\omega_{\mathrm{C}}-\omega_0\right)t\right)+\mathrm{Si}\left(\omega_0 t\right)\right)\vphantom{\frac{1}{\omega_{\mathrm{C}}-\omega_0}}\right]
\end{multline}
where $\mathrm{Ci}$ stands for the cosine integral and $\mathrm{Si}$ for the sine integral \cite{AbramowitzStegun}, is strongly dependent on the value of the cutoff frequency\footnote{We also note that when the interaction Hamiltonian is taken to be of the $\hat{\mathbf{E}}\cdot\hat{\mathbf{x}}$ form instead of the $\hat{\mathbf{A}}\cdot\hat{\mathbf{p}}$ form, the divergence of the integral corresponding to (\ref{eq:NoChanceSurvival}), which features $\omega^3$ instead of $\omega$, is quadratic instead of logarithmic, and the dependence of the truncated integral (\ref{eq:Trunk}) -which features the same substitution- on the cutoff frequency is very much enhanced. See \cite{EdouardArXiv}. \label{ftn:Test}} $\omega_{\mathrm{C}}$. A discussion of the relevance of (\ref{eq:Trunk}) is made in sect.~\ref{sec:Ccl}. In a recent paper \cite{Norge} devoted to the decay of magnetic dipoles, similar questions were raised, and the Compton frequency was proposed as a cutoff. With this cutoff, Grimsmo \emph{et al.} proposed a regularisation of their problem following the lines of Bethe's mass renormalisation.\\}
\item{When a cutoff is implemented, the distinction between electromagnetic modes which are considered to be coupled to the atom and those who are excluded from the treatment is binary: the coupling function is taken to be exactly zero beyond the cutoff frequency. One could envision to introduce a smoother cutoff procedure along the lines of the cutting off of ultrarelativistic frequencies presented in \cite{CohenQED1}, but we feel this would do nothing but introduce further arbitrariness in the model.}
\end{itemize}

\subsection{The dipole approximation: regularisation} \label{subsec:DipoleCalc}

For the reasons we gave in the previous sect.~\ref{subsec:DipoleDsc}, we follow a different path: we refrain from introducing a cutoff and instead retain the result (\ref{eq:NoChanceSurvival}) from the dipole approximation without cutoff, but will endeavour to regularise the divergence in the integral. We will then extract the regular (finite) terms, and inspect them carefully. As far as we know this treatment of the present problem is novel.\\

We focus our interest on the integral featured in (\ref{eq:NoChanceSurvival}), that is
\begin{multline} \label{eq:Frederic}
\int_{-\infty}^{+\infty}\mathrm{d}\omega\,f\left(\omega,t\right)\hspace{15pt}\text{where}\\
f\left(\omega,t\right)=\theta\left(\omega\right)\omega\sin\!\mathrm{c}^2\left(\left(\omega_0-\omega\right)\frac{t}{2}\right)\equiv\theta\left(\omega\right)g\left(\omega,t\right).
\end{multline}
Here $\theta$ stands for the Heaviside step distribution. As we have discussed at length, (\ref{eq:Frederic}) is a divergent integral. Nevertheless, we shall extract its regular (finite) terms.\\ 

The idea here is that although the function $f\left(\cdot,t\right)$ does not belong to the vector space $L^1\left(\mathbb{R}\right)$ of summable functions, it is a slowly growing function, and, as such, a tempered distribution. It therefore admits a Fourier transform in the sense of distributions, which we write $\bar{f}\left(\cdot,t\right)$.\\

We thus compute the Fourier transform
\begin{equation} \label{eq:DistFourier}
\bar{f}\left(\tau,t\right)=\int_{-\infty}^{+\infty}\mathrm{d}\omega\,f\left(\omega,t\right)\mathrm{e}^{-\mathrm{i}\omega\tau}
\end{equation}
and then shall take the limit $\tau\rightarrow0$ at the end to retrieve the desired integral (\ref{eq:Frederic}). In this limit, some terms in $\bar{f}\left(\tau,t\right)$ become ill-defined, a consequence of the fact that $f\left(\cdot,t\right)$ is not summable. We will simply discard these terms at the end of our treatment, and focus on the well-defined terms in the limit $\tau\rightarrow0$.\\

The folding theorem and the well-known expression for the Fourier transform of the Heaviside distribution yield, from (\ref{eq:Frederic}) and (\ref{eq:DistFourier})
\begin{equation} \label{eq:ConvWithDeltavp}
\bar{f}\left(\tau,t\right)=\frac{1}{2}\left[\left(\delta\left(\cdot\right)-\frac{\mathrm{i}}{\pi}\mathrm{vp}\,\frac{1}{\cdot}\right)*\bar{g}\left(\cdot,t\right)\right]\left(\tau\right)
\end{equation}
where $\mathrm{vp}$ stands for the Cauchy principal value of the subsequent function and the relation between $\bar{g}$ and $g$ is the same as that (\ref{eq:DistFourier}) between $\bar{f}$ and $f$:
\begin{equation} \label{eq:DistFourierAgain}
\bar{g}\left(\tau,t\right)=\int_{-\infty}^{+\infty}\mathrm{d}\omega\,g\left(\omega,t\right)\mathrm{e}^{-\mathrm{i}\omega\tau}.
\end{equation}
From (\ref{eq:Frederic}) we get
\begin{multline} \label{eq:ExpandSine}
\bar{g}\left(\tau,t\right)=-\frac{1}{t^2}\int_{-\infty}^{+\infty}\mathrm{d}\omega\frac{\omega}{\left(\omega-\omega_0\right)^2}\\\left(\mathrm{e}^{\mathrm{i}\omega\left(t-\tau\right)}\mathrm{e}^{-\mathrm{i}\omega_0t}+\mathrm{e}^{-\mathrm{i}\omega\left(t+\tau\right)}\mathrm{e}^{\mathrm{i}\omega_0t}-2\mathrm{e}^{-\mathrm{i}\omega\tau}\right).
\end{multline}
We would like to compute this as a sum of three integrals corresponding to the three summands on the right-hand side of (\ref{eq:ExpandSine}), but, taken individually, these integrals will diverge because of the singularity at $\omega=\omega_0$. The full integrand in (\ref{eq:ExpandSine}), however, has no singularity at $\omega=\omega_0$. Accordingly we introduce a small positive imaginary part $\epsilon>0$ in the denominator, which enables us to compute (\ref{eq:ExpandSine}) as a sum of three integrals. Introduce
\begin{equation} \label{eq:CapitalG}
G_\epsilon\left(\tau\right)\equiv\int_{-\infty}^{+\infty}\mathrm{d}\omega\frac{\omega}{\left(\omega-\omega_0+\mathrm{i}\epsilon\right)^2}\mathrm{e}^{-\mathrm{i}\omega\tau}
\end{equation}
to rewrite
\begin{equation} \label{eq:UppertoLower}
\bar{g}\left(\tau,t\right)=-\frac{1}{t^2}\lim_{\epsilon\to0^+}\left(\mathrm{e}^{-\mathrm{i}\omega_0t}G_\epsilon\left(\tau-t\right)+\mathrm{e}^{\mathrm{i}\omega_0t}G_\epsilon\left(\tau+t\right)-2G_\epsilon\left(\tau\right)\right).
\end{equation}
Thus to compute $\bar{g}\left(\cdot,t\right)$ we need only compute $G_\epsilon$, which we do now. Notice first that
\begin{equation} \label{eq:Apart}
\frac{\omega}{\left(\omega-\omega_0+\mathrm{i}\epsilon\right)^2}=\frac{1}{\omega-\omega_0+\mathrm{i}\epsilon}+\frac{\omega_0+\mathrm{i}\epsilon}{\left(\omega-\omega_0+\mathrm{i}\epsilon\right)^2}.
\end{equation}
Since $\epsilon>0$, an application of Cauchy's residue theorem therefore yields
\begin{equation} \label{eq:ThanksAugustin}
\lim_{\epsilon\to0^+}G_\epsilon\left(\tau\right)=2\pi\mathrm{e}^{-\mathrm{i}\omega_0\tau}\left(\mathrm{i}+\omega_0\tau\right)\theta\left(\tau\right).
\end{equation}
Plugging this back in (\ref{eq:UppertoLower}), we get
\begin{multline} \label{eq:ProvisionalEnd}
\bar{g}\left(\tau,t\right)=2\pi\frac{1}{t^2}\mathrm{e}^{-\mathrm{i}\omega_0\tau}\left[\theta\left(\tau-t\right)\left(\mathrm{i}+\omega_0\left(\tau-t\right)\right)\right.\\\left.+\theta\left(\tau+t\right)\left(\mathrm{i}+\omega_0\left(\tau+t\right)\right)-2\theta\left(\tau\right)\left(\mathrm{i}+\omega_0\tau\right)\right].
\end{multline}
Then we further plug this in (\ref{eq:ConvWithDeltavp}) to get
\begin{multline} \label{eq:DeltaPlusvp}
\hspace{-12.5pt}\bar{f}\left(\tau,t\right)=\frac{1}{2}\bar{g}\left(\tau,t\right)-\frac{\mathrm{i}}{t^2}\mathrm{vp}\int_{-\infty}^{+\infty}\mathrm{d}\sigma\frac{\mathrm{e}^{-\mathrm{i}\omega_0\sigma}}{\tau-\sigma}\left[\theta\left(\sigma-t\right)\left(\mathrm{i}+\omega_0\left(\sigma-t\right)\right)\right.\\\left.+\theta\left(\sigma+t\right)\left(\mathrm{i}+\omega_0\left(\sigma+t\right)\right)-2\theta\left(\sigma\right)\left(\mathrm{i}+\omega_0\sigma\right)\right].
\end{multline}
After some algebra we obtain
\begin{multline} \label{eq:Finalfortau}
\bar{f}\left(\tau,t\right)=\frac{1}{t^2}\left\{\pi\left[\theta\left(\tau-t\right)\left(\mathrm{i}+\omega_0\left(\tau-t\right)\right)\right.\right.\\\left.\left.+\theta\left(\tau+t\right)\left(\mathrm{i}+\omega_0\left(\tau+t\right)\right)\right.\right.\\\left.\left.-2\theta\left(\tau\right)\left(\mathrm{i}+\omega_0\tau\right)\right]-\mathrm{i}\,\mathrm{vp}\left[\mathrm{i}\left(\int_{-t}^0\mathrm{d}\sigma\frac{\mathrm{e}^{-\mathrm{i}\omega_0\sigma}}{\tau-\sigma}-\int_0^t\mathrm{d}\sigma\frac{\mathrm{e}^{-\mathrm{i}\omega_0\sigma}}{\tau-\sigma}\right)\right.\right.\\\left.\left.+\omega_0\left(t\int_{-t}^t\mathrm{d}\sigma\frac{\mathrm{e}^{-\mathrm{i}\omega_0\sigma}}{\tau-\sigma}+\int_{-t}^0\mathrm{d}\sigma\frac{\mathrm{e}^{-\mathrm{i}\omega_0\sigma}}{\tau-\sigma}\sigma-\int_0^t\mathrm{d}\sigma\frac{\mathrm{e}^{-\mathrm{i}\omega_0\sigma}}{\tau-\sigma}\sigma\right)\right]\right\}.
\end{multline}
Hence we can finally write
\begin{multline} \label{eq:RegSng}
\bar{f}\left(\tau=0,t\right)=\frac{1}{t^2}\left\{\pi\left[\theta\left(-t\right)\left(\mathrm{i}-\omega_0t\right)+\theta\left(t\right)\left(\mathrm{i}+\omega_0t\right)\right.\right.\\\left.\left.-2\mathrm{i}\theta\left(0\right)\right]-\mathrm{i}\,\mathrm{vp}\left[\mathrm{i}\left(\int_{-t}^0\mathrm{d}\sigma\frac{\mathrm{e}^{-\mathrm{i}\omega_0\sigma}}{-\sigma}-\int_0^t\mathrm{d}\sigma\frac{\mathrm{e}^{-\mathrm{i}\omega_0\sigma}}{-\sigma}\right)\right.\right.\\\left.\left.+\omega_0\left(t\int_{-t}^t\mathrm{d}\sigma\frac{\mathrm{e}^{-\mathrm{i}\omega_0\sigma}}{-\sigma}-\int_{-t}^0\mathrm{d}\sigma\,\mathrm{e}^{-\mathrm{i}\omega_0\sigma}+\int_0^t\mathrm{d}\sigma\,\mathrm{e}^{-\mathrm{i}\omega_0\sigma}\right)\right]\right\}.
\end{multline}
Now we want to identify and discard the singular terms in (\ref{eq:RegSng}). Making use of $\theta\left(-t\right)+\theta\left(t\right)=1$, we can rewrite $2\theta\left(0\right)=1+\theta\left(0\right)-\theta\left(-0\right)$. The quantity $\theta\left(0\right)-\theta\left(-0\right)$ is singular\footnote{It can be understood as equal to $\mathrm{sgn}\left(0\right)$.}, and we discard it. We claim that the difference of twe two (principal value) integrals on the second line on the right-hand side of (\ref{eq:RegSng}) also features a singular term. Let us show this. Rewrite
\begin{equation} \label{eq:ExtractSng}
\begin{aligned} [b]
\int_{-t}^0\mathrm{d}\sigma\frac{\mathrm{e}^{-\mathrm{i}\omega_0\sigma}}{-\sigma}-\int_0^t\mathrm{d}\sigma\frac{\mathrm{e}^{-\mathrm{i}\omega_0\sigma}}{-\sigma}&=2\int_0^t\mathrm{d}\sigma\frac{\cos\left(\omega_0\sigma\right)}{\sigma}\\
&=2\sum_{n=0}^{+\infty}\frac{\left(-1\right)^n}{\left(2n\right)!}\omega_0^{2n}\int_0^t\mathrm{d}\sigma\,\sigma^{2n-1}.
\end{aligned}
\end{equation}
Now, for the $n=0$ term in this sum, the integral diverges, and taking its principal value will not change that fact. Accordingly, we simply discard the $n=0$ term in (\ref{eq:ExtractSng}). We write the remainder of the series in closed form:
\begin{equation} \label{eq:CosineInt}
\begin{aligned} [b]
\sum_{n=1}^{+\infty}\frac{\left(-1\right)^n}{\left(2n\right)!}\omega_0^{2n}\int_0^t\mathrm{d}\sigma\,\sigma^{2n-1}&=\sum_{n=1}^{+\infty}\frac{\left(-1\right)^n}{2n\left(2n\right)!}\left(\omega_0t\right)^{2n}\\
&=\mathrm{Ci}\left(\omega_0t\right)-\log\left(\omega_0t\right)-\gamma
\end{aligned}
\end{equation}
where $\gamma$ is the Euler-Mascheroni constant. Computing the integrals on the third line of (\ref{eq:RegSng}), we can rewrite the regular part of (\ref{eq:RegSng}) as
\begin{multline} \label{eq:RegOnly}
\bar{f}\left(\tau=0,t\right)\stackrel{\mathrm{r.p.}}{=}\frac{1}{t^2}\left[-4\sin^2\left(\frac{\omega_0 t}{2}\right)+2\left(\mathrm{Ci}\left(\omega_0t\right)-\log\left(\omega_0t\right)-\gamma\right)\right.\\\left.+\pi\omega_0 t\left(\mathrm{sgn}\left(t\right)+\frac{2}{\pi}\mathrm{Si}\left(\omega_0 t\right)\right)\right]
\end{multline}
where $\mathrm{r.p.}$ stands for ``regular part'. The term
\begin{equation} \label{eq:FermiSeed}
\pi\frac{\omega_0}{t}\left(\mathrm{sgn}\left(t\right)+\frac{2}{\pi}\mathrm{Si}\left(\omega_0 t\right)\right)
\end{equation}
on the right-hand side of (\ref{eq:RegOnly}) is particularly interesting, and can be directly linked to Fermi's golden rule. Indeed, remember from (\ref{eq:NoChanceSurvival}) that the decay probability (that is, $1-P_{\mathrm{surv}}\left(t\right)$) features the product of (\ref{eq:RegOnly}) by $t^2$. Further, notice that $\mathrm{Si}\left(\omega_0 t\right)$ quickly converges to the Dirichlet value $\pi/2$ as $t$ becomes substantially larger than $1/\omega_0$. For such times the leading term in (\ref{eq:RegOnly}) is clearly $2\pi\omega_0/t$ (\emph{i.e.}, the limit of (\ref{eq:FermiSeed}) as $t\rightarrow+\infty$) which is equal to the result obtained from illegally ``sneaking'' the limit $t\rightarrow+\infty$ into the divergent integral on the right-hand side of (\ref{eq:NoChanceSurvival}). Hence, we have shown how the golden rule can be retrieved from a formal, cutoff-independent regularisation of the integral featured in the expression for the survival probability. We refrain from claiming that the terms on the first line of the right-hand side of (\ref{eq:RegOnly}) are relevant descriptions of short-time deviations from the golden rule, as it is clear that using a more exact expression for the atom-field coupling will give better results\footnote{For the sake of exhaustiveness, the regularised dipole-approximated result (\ref{eq:RegOnly}) is plotted in Figs.~\ref{fig:AlmostFermi} and \ref{fig:ZenoFermi}, where it is shown that it does not provide an accurate description of the very short-time behaviour of the system.}. This is examined in the upcoming sect.~\ref{subsec:MultipoleCalc}. Nevertheless, we can notice that (\ref{eq:RegOnly}) tends to zero as $t\rightarrow0$, which is an agreeable feature of our result.

%\footnote{We initially studied (as mentioned in footnote~\ref{ftn:Test} page \pageref{ftn:Test}) this problem with the interaction Hamiltonian $\hat{\mathbf{E}}\cdot\hat{\mathbf{x}}$ instead of $\hat{\mathbf{A}}\cdot\hat{\mathbf{p}}$. In this case, several extra singular terms appear in the sister expression to (\ref{eq:RegSng}). When one extracts the regular terms, one obtains an expresion very similar to (\ref{eq:RegOnly}), with two extra terms that feature, respectively, the Cauchy principal value of $1/t$ and the Hadamard principal part of $1/t^2$. DISCUSS (ANDRÉ).}

\subsection{Exact coupling: vindication of Fermi's golden rule} \label{subsec:MultipoleCalc}

In the present section we will start from the same integral (\ref{eq:CardSineSurv}), and use the exact matrix element (\ref{eq:MatrixMultipole}). This treatment features no infinites and thus does not call for any regularisation procedure. Plus, it allows us to investigate short time deviations from Fermi's golden rule in a more direct and reliable way. Start from the survival probability of the electron in the excited state
\begin{multline} \label{eq:SomeChanceSurvival}
P_{\mathrm{surv}}\left(t\right)=1-\frac{2^{10}}{3^{9}\pi}c^2\,\alpha^3\,t^2\\\int_0^{+\infty}\mathrm{d}\omega\,\frac{\omega}{\left[1+\left(\frac{\omega}{\omega_\mathrm{X}}\right)^2\right]^4}\sin\!\mathrm{c}^2\left(\left(\omega_0-\omega\right)\frac{t}{2}\right)
\end{multline}
where we indroduced the notation $\omega_{\mathrm{X}}\equiv\left(3/2\right)\left(c/a_0\right)$. The frequency $\omega_{\mathrm{X}}$ is a natural cutoff frequency coming from the exact computation of the interaction matrix element (\ref{eq:MatrixMultipole}). The integral in (\ref{eq:SomeChanceSurvival}) is finite at all times and we can compute it numerically or, as we shall now see, analytically.\\

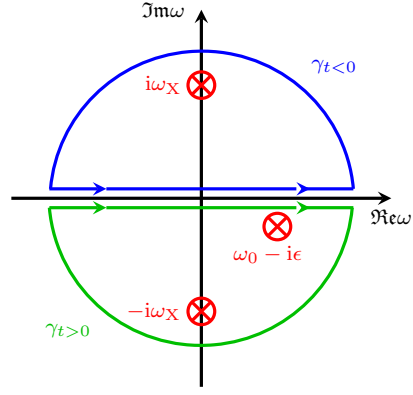
\begin{figure} [t]
\begin{center}
\begin{tikzpicture}[very thick, >=stealth]
\draw[->] (-2.5,0) -- (2.5,0);
\draw[->] (0,-2.5) -- (0,2.5);
\draw (2.5,-.25) node {\small $\mathfrak{Re}\omega$};
\draw (-.5,2.5) node {\small $\mathfrak{Im}\omega$};
\draw[red] (.875,-.5) -- (1.125,-.25);
\draw[red] (.875,-.25) -- (1.125,-.5);
\draw[red] (1,-.375) circle (5pt);
\draw[red] (.875,-.75) node {\small $\omega_0-\mathrm{i}\epsilon$};
\draw[red] (-.125,1.625) -- (.125,1.375);
\draw[red] (-.125,1.375) -- (.125,1.625);
\draw[red] (0,1.5) circle (5pt);
\draw[red] (-.5,1.5) node {\small $\mathrm{i}\omega_{\mathrm{X}}$};
\draw[red] (-.125,-1.625) -- (.125,-1.375);
\draw[red] (-.125,-1.375) -- (.125,-1.625);
\draw[red] (0,-1.5) circle (5pt);
\draw[red] (-.625,-1.5) node {\small $-\mathrm{i}\omega_{\mathrm{X}}$};
\draw[blue, ->] (-2,.125) -- (-1.25,.125);
\draw[blue] (-1.25,.125) -- (1.25,.125);
\draw[blue, >-] (1.25,.125) -- (2,.125);
\draw[blue] (2,.125) arc (5:175:2);
\draw[blue] (1.75,1.75) node {\small $\gamma_{t<0}$};
\draw[green!75!black, -<] (2,-.125) -- (1.25,-.125);
\draw[green!75!black] (1.25,-.125) -- (-1.25,-.125);
\draw[green!75!black, <-] (-1.25,-.125) -- (-2,-.125);
\draw[green!75!black] (-2,-.125) arc (185:355:2);
\draw[green!75!black] (-1.75,-1.75) node {\small $\gamma_{t>0}$};
\end{tikzpicture}
\end{center}
\vspace{-10pt}
\caption{Jordan loops in the complex $k$-plane used to compute the Fourier transform (\ref{eq:TrueStart}). The poles $\omega_0-\mathrm{i}\epsilon$ and $\pm\mathrm{i}\omega_{\mathrm{X}}$ of the integrand are represented by red circled crosses. \label{fig:CauchyPaths}}
\end{figure}

Define
\begin{multline} \label{eq:FermiIntegral}
I_F\left(t\right)=\int_{-\infty}^{+\infty}\mathrm{d}\omega\,d\left(\omega,t\right)\hspace{15pt}\text{where}\\
\hspace{-5pt}d\left(\omega,t\right)=\theta\left(\omega\right)\frac{\omega}{\left[1+\left(\frac{\omega}{\omega_\mathrm{X}}\right)^2\right]^4}\sin\!\mathrm{c}^2\left(\left(\omega_0-\omega\right)\frac{t}{2}\right).
\end{multline}
It can be rewritten
\begin{multline} \label{eq:ExpandSineAgain}
I_F\left(t\right)=-\frac{1}{t^2}\int_0^{+\infty}\mathrm{d}\omega\frac{\omega}{\left(\omega-\omega_0\right)^2}\frac{1}{\left[1+\left(\frac{\omega}{\omega_\mathrm{X}}\right)^2\right]^4}\\\left(\mathrm{e}^{\mathrm{i}\omega t}\mathrm{e}^{-\mathrm{i}\omega_0t}+\mathrm{e}^{-\mathrm{i}\omega t}\mathrm{e}^{\mathrm{i}\omega_0t}-2\right).
\end{multline}
We would like to compute this as a sum of three integrals corresponding to the three summands on the right-hand side of (\ref{eq:ExpandSineAgain}), but, taken individually, these integrals will diverge because of the singularity at $\omega=\omega_0$. The full integrand in (\ref{eq:ExpandSineAgain}), however, has no singularity at $\omega=\omega_0$. Accordingly we introduce a small positive imaginary part $\epsilon>0$ in the denominator. Introduce
\begin{subequations} \label{eq:CapitalH}
\begin{align}
H_\epsilon\left(t\right)&\equiv\int_{-\infty}^{+\infty}\mathrm{d}\omega\frac{\omega}{\left(\omega-\omega_0+\mathrm{i}\epsilon\right)^2}\frac{\theta\left(\omega\right)}{\left[1+\left(\frac{\omega}{\omega_\mathrm{X}}\right)^2\right]^4}\mathrm{e}^{-\mathrm{i}\omega t}\\
&\equiv\int_{-\infty}^{+\infty}\mathrm{d}\omega\,h_\epsilon\left(\omega\right)\theta\left(\omega\right)\mathrm{e}^{-\mathrm{i}\omega t}
\end{align}
\end{subequations}
to rewrite
\begin{equation} \label{eq:Target}
I_F\left(t\right)=-\frac{1}{t^2}\lim_{\epsilon\to0^+}\left(\mathrm{e}^{-\mathrm{i}\omega_0t}H_\epsilon\left(-t\right)+\mathrm{e}^{\mathrm{i}\omega_0t}H_\epsilon\left(t\right)-2H_\epsilon\left(0\right)\right).
\end{equation}
Thus to compute $I_F$ we need only compute $H_\epsilon$, which we do now. The folding theorem and the well-known expression for the Fourier transform of the Heaviside distribution yield
\begin{equation} \label{eq:ConvWithDeltavpAgain}
H_\epsilon\left(t\right)=\frac{1}{2}\left[\left(\delta\left(\cdot\right)-\frac{\mathrm{i}}{\pi}\mathrm{vp}\,\frac{1}{\cdot}\right)*\bar{h}_\epsilon\left(\cdot\right)\right]\left(t\right).
\end{equation}
We therefore need to compute the Fourier transform
\begin{equation} \label{eq:TrueStart}
\bar{h}_\epsilon\left(t\right)=\int_{-\infty}^{+\infty}\mathrm{d}\omega\,h_\epsilon\left(\omega\right)\mathrm{e}^{-\mathrm{i}\omega t}
\end{equation}
of $h_\epsilon$. We use Cauchy's residue theorem. We know from (\ref{eq:CapitalH}) that $h_\epsilon$ has a second order pole at $\omega_0-\mathrm{i}\epsilon$ and two fourth order poles at $\pm\mathrm{i}\omega_{\mathrm{X}}$, pictured on Fig.~\ref{fig:CauchyPaths}. From (\ref{eq:TrueStart}) we see that we have to close the integration path (Jordan loop) in the lower half of the complex plane for $t>0$, and in the upper half of the plane for $t<0$.\\

It can be checked that the residues of $h_\epsilon\left(\omega\right)\mathrm{e}^{-\mathrm{i}\omega t}$ read
\begin{subequations} \label{eq:Residues}
\begin{align}
\mathrm{Res}\left(h_\epsilon\left(\cdot\right)\mathrm{e}^{-\mathrm{i}\cdot t},\omega_0-\mathrm{i}\epsilon\right)&=\mathrm{e}^{-\mathrm{i}\left(\omega_0-\mathrm{i}\epsilon\right)t}\left(a_0+a_1\,t\right),\\
\mathrm{Res}\left(h_\epsilon\left(\cdot\right)\mathrm{e}^{-\mathrm{i}\cdot t},\mathrm{i}\omega_{\mathrm{X}}\right)&=\mathrm{e}^{\omega_{\mathrm{X}}t}\left(b_0^++b_1^+\,t+b_2^+\,t^2+b_3^+\,t^3\right),\\
\mathrm{Res}\left(h_\epsilon\left(\cdot\right)\mathrm{e}^{-\mathrm{i}\cdot t},-\mathrm{i}\omega_{\mathrm{X}}\right)&=\mathrm{e}^{-\omega_{\mathrm{X}}t}\left(b_0^-+b_1^-\,t+b_2^-\,t^2+b_3^-\,t^3\right)\end{align}
\end{subequations}
where the $a_i$ and $b_i^\pm$ coefficients depend on $\epsilon$. Whence the Fourier transform (\ref{eq:TrueStart})
\begin{multline} \label{eq:CauchyResidue}
\bar{h}_\epsilon\left(t\right)=-2\mathrm{i}\pi\left[\theta\left(t\right)\mathrm{e}^{-\mathrm{i}\left(\omega_0-\mathrm{i}\epsilon\right)t}\left(a_0+a_1\,t\right)\right.\\\left.+\theta\left(t\right)\mathrm{e}^{-\omega_{\mathrm{X}}t}\left(b_0^++b_1^+\,t+b_2^+\,t^2+b_3^+\,t^3\right)\right.\\\left.-\theta\left(-t\right)\mathrm{e}^{\omega_{\mathrm{X}}t}\left(b_0^-+b_1^-\,t+b_2^-\,t^2+b_3^-\,t^3\right)\right].
\end{multline}
One can then deduce $H_\epsilon\left(t\right)$ and its limit as $\epsilon\rightarrow0^+$. Setting
\begin{subequations} \label{eq:LowertoUpper}
\begin{align}
a_i&\underset{\epsilon\rightarrow0^+}{\longrightarrow}A_i,\\
b_i^\pm&\underset{\epsilon\rightarrow0^+}{\longrightarrow}B_i^\pm,\\
\end{align}
\end{subequations}
one can see that
\begin{align*}
B_0^+=B_0^{-*}&\equiv B_0,\\
B_1^+=-B_1^{-*}&\equiv B_1,\\
B_2^+=B_2^{-*}&\equiv B_2,\\
B_3^+=-B_3^{-*}&\equiv B_3.
\end{align*}
We give
\begin{subequations} \label{eq:CapitalAGalore}
\begin{align}
A_0&=\frac{\omega_{\mathrm{X}}^8 \left(\omega_{\mathrm{X}}^2-7\omega_0^2\right)}{\left(\omega_0^2+\omega_{\mathrm{X}}^2\right)^5},\\
A_1&=-\mathrm{i}\frac{\omega_0}{\left[1+\left(\frac{\omega_0}{\omega_{\mathrm{X}}}\right)^2\right]^4}\end{align}
\end{subequations}
and
\begin{subequations} \label{eq:CapitalBGalore}
\begin{align}
B_0&=-\frac{\omega_{\mathrm{X}}^3 \left(-6 \omega_0^2+30\mathrm{i}\omega_0 \omega_{\mathrm{X}}+48 \omega_{\mathrm{X}}^2\right)}{96 (\omega_{\mathrm{X}}+\mathrm{i}\omega_0)^5},\\
B_1&=\frac{\omega_{\mathrm{X}}^3 \left(-3\mathrm{i}\omega_0^3-21 \omega_0^2 \omega_{\mathrm{X}}+51\mathrm{i}\omega_0 \omega_{\mathrm{X}}^2+33 \omega_{\mathrm{X}}^3\right)}{96 (\omega_{\mathrm{X}}+\mathrm{i}\omega_0)^5},\\
B_2&=-\frac{\omega_{\mathrm{X}}^3 \left(-3\mathrm{i}\omega_0^3 \omega_{\mathrm{X}}-15 \omega_0^2 \omega_{\mathrm{X}}^2+21\mathrm{i}\omega_0 \omega_{\mathrm{X}}^3+9 \omega_{\mathrm{X}}^4\right)}{96 (\omega_{\mathrm{X}}+\mathrm{i}\omega_0)^5},\\
B_3&=\frac{\omega_{\mathrm{X}}^3 \left(-\mathrm{i} \omega_0^3 \omega_{\mathrm{X}}^2-3 \omega_0^2 \omega_{\mathrm{X}}^3+3\mathrm{i}\omega_0 \omega_{\mathrm{X}}^4+\omega_{\mathrm{X}}^5\right)}{96 (\omega_{\mathrm{X}}+\mathrm{i}\omega_0)^5}.
\end{align}
\end{subequations}
The computation of $H_{0^+}\left(t\right)$ from (\ref{eq:ConvWithDeltavpAgain}) and (\ref{eq:CauchyResidue}) features no notable conceptual or technical difficulty, but is quite tedious. We only give the result, which reads
\begin{multline} \label{eq:PostHardcore}
H_{0^+}\left(t\right)=-\mathrm{i}\pi\left[\theta\left(t\right)\mathrm{e}^{-\mathrm{i}\omega_0t}\left(A_0+A_1\,t\right)\right.\\\left.+\theta\left(t\right)\mathrm{e}^{-\omega_{\mathrm{X}}t}\left(B_0^++B_1^+\,t+B_2^+\,t^2+B_3^+\,t^3\right)\right.\\\left.-\theta\left(-t\right)\mathrm{e}^{\omega_{\mathrm{X}}t}\left(B_0^-+B_1^-\,t+B_2^-\,t^2+B_3^-\,t^3\right)\vphantom{\mathrm{e}^{\mathrm{i}}}\right]\\+\mathrm{e}^{-\mathrm{i}\omega_0 t}\left[-\mathrm{i}\left(A_0+A_1\,t\right)\mathrm{Si}\left(\omega_0t\right)+\frac{2}{\omega_0}A_1\sin\left(\omega_0t\right)\right.\\\left.-\left(A_0+A_1\,t\right)\left(\mathrm{Ci}\left(\omega_0t\right)+\mathrm{i}\frac{\pi}{2}\right)-\frac{\mathrm{i}}{\omega_0}A_1\mathrm{e}^{\mathrm{i}\omega_0t}\right]\\-\left[\left(B_0^*-B_1^*\,t+B_2^*\,t^2-B_3^*\,t^3\right)\mathrm{e}^{-\omega_{\mathrm{X}}t}\mathrm{Ei}\left(\omega_{\mathrm{X}}t\right)\right.\\\left.+\left(B_0+B_1\,t+B_2\,t^2+B_3\,t^3\right)\mathrm{e}^{\omega_{\mathrm{X}}t}\mathrm{Ei}\left(-\omega_{\mathrm{X}}t\right)\right]\\-\frac{1}{\omega_{\mathrm{X}}}\left[\left(B_1+B_1^*\right)-\left(B_2+B_2^*\right)t+2\left(B_3+B_3^*\right)t^2\right]\\-\frac{1}{\omega_{\mathrm{X}^2}}\left[\left(B_2-B_2^*\right)\omega_{\mathrm{X}}t+2\left(B_3-B_3^*\right)t\right]-\frac{1}{\omega_{\mathrm{X}^3}}\left(B_3+B_3^*\right)\omega_{\mathrm{X}}^2t^2.
\end{multline}
Here $\mathrm{Ei}$ stands for the exponential integral \cite{AbramowitzStegun}. Now remains the unenviable task of adding three such terms as prescribed by (\ref{eq:Target}), so as to obtain the exact expression for the survival probability as given by first-order time-dependent perturbation theory. Simplifications are frankly scarce here, but we are ``saved'' from the apparition of any singular terms by the identity
\begin{equation} \label{eq:SavingGrace}
A_0+B_0+B_0^*=0.
\end{equation}
The link between (\ref{eq:SavingGrace}) and the presence/absence of singular terms is explained as follows. As seen from (\ref{eq:Residues}), the quantity $A_0+B_0+B_0^*$ is, up to a factor of $\pm2\mathrm{i}\pi$, equal to the integral of $h_{0^+}$ over any closed curve $\Gamma$ circling around the three poles of $h_{0^+}$ (see Fig.~\ref{fig:CauchyPaths}). We can take $\Gamma$ to be a circle of radius $R$ centred around $z=0$. It is easy to see from Jordan's lemmas that the integral of $h_{0^+}$ over such a curve vanishes when $R\rightarrow+\infty$\footnote{For large $\omega$, it is easy to see that $h_{0^+}$ behaves as $\omega^{-9}$.} (and hence for any $R$ large enough that the circle will still enclose the three poles), whence (\ref{eq:SavingGrace}).\\

With the help of (\ref{eq:SavingGrace}) we can finally write, from (\ref{eq:Target}) and (\ref{eq:PostHardcore}),
{\allowdisplaybreaks
\begin{multline} \label{eq:IntheEnd}
I_F\left(t\right)=\frac{1}{t^2}\left\{-2A_0\left(\log\left(\frac{\omega_0}{\omega_{\mathrm{X}}}\right)-\mathrm{Ci}\left(\omega_0\left|t\right|\right)\right)+\mathrm{i}\pi\left(B_0-B_0^*\right)\right.\\\left.+A_1\left[-4\frac{\mathrm{i}}{\omega_0}\sin^2\left(\frac{\omega_0t}{2}\right)+\mathrm{i}\pi t\left(\mathrm{sgn}\left(t\right)+\frac{2}{\pi}\mathrm{Si}\left(\omega_0t\right)\right)\right]\right.\\\left.+\mathrm{i}\pi\left[\mathrm{e}^{-\omega_{\mathrm{X}}t}\theta\left(t\right)\left[\left(B_0^*\,\mathrm{e}^{\mathrm{i}\omega_0t}-B_0\,\mathrm{e}^{-\mathrm{i}\omega_0t}\right)\right.\right.\right.\\\left.\left.\left.+\left(-B_1^*\,\mathrm{e}^{\mathrm{i}\omega_0t}+B_1\,\mathrm{e}^{-\mathrm{i}\omega_0t}\right)t\right.\right.\right.\\\left.\left.\left.+\left(B_2^*\,\mathrm{e}^{\mathrm{i}\omega_0t}-B_2\,\mathrm{e}^{-\mathrm{i}\omega_0t}\right)t^2+\left(-B_3^*\,\mathrm{e}^{\mathrm{i}\omega_0t}+B_3\,\mathrm{e}^{-\mathrm{i}\omega_0t}\right)t^3\right)\right.\right.\\\left.\left.+\mathrm{e}^{\omega_{\mathrm{X}}t}\theta\left(-t\right)\left(\left(B_0^*\,\mathrm{e}^{-\mathrm{i}\omega_0t}-B_0\,\mathrm{e}^{\mathrm{i}\omega_0t}\right)\right.\right.\right.\\\left.\left.\left.-\left(-B_1^*\,\mathrm{e}^{-\mathrm{i}\omega_0t}+B_1\,\mathrm{e}^{\mathrm{i}\omega_0t}\right)t\right.\right.\right.\\\left.\left.\left.+\left(B_2^*\,\mathrm{e}^{-\mathrm{i}\omega_0t}-B_2\,\mathrm{e}^{\mathrm{i}\omega_0t}\right)t^2-\left(-B_3^*\,\mathrm{e}^{-\mathrm{i}\omega_0t}+B_3\,\mathrm{e}^{\mathrm{i}\omega_0t}\right)t^3\right)\right]\right.\\\left.+\mathrm{e}^{-\omega_{\mathrm{X}}t}\mathrm{Ei}\left(\omega_{\mathrm{X}}t\right)\left(\left(B_0^*-B_1^*\,t+B_2^*\,t^2-B_3^*\,t^3\right)\mathrm{e}^{\mathrm{i}\omega_0t}\right.\right.\\\left.\left.+\left(B_0-B_1\,t+B_2\,t^2-B_3\,t^3\right)\mathrm{e}^{-\mathrm{i}\omega_0t}\right)\right.\\\left.+\mathrm{e}^{\omega_{\mathrm{X}}t}\mathrm{Ei}\left(-\omega_{\mathrm{X}}t\right)\left(\left(B_0^*+B_1^*\,t+B_2^*\,t^2+B_3^*\,t^3\right)\mathrm{e}^{-\mathrm{i}\omega_0t}\right.\right.\\\left.\left.+\left(B_0+B_1\,t+B_2\,t^2+B_3\,t^3\right)\mathrm{e}^{\mathrm{i}\omega_0t}\right)\right.\\\left.+\frac{2}{\omega_{\mathrm{X}}^3}\left[-2\left(\left(B_1+B_1^*\right)\omega_{\mathrm{X}}^2-\left(B_2+B_2^*\right)\omega_{\mathrm{X}}\right.\right.\right.\\\left.\left.\left.+2\left(B_3+B_3^*\right)\vphantom{\omega_{\mathrm{X}}^2}\right)\sin^2\left(\frac{\omega_0t}{2}\right)\right.\right.\\\left.\left.+\mathrm{i}\left(\left(B_2-B_2^*\right)\omega_{\mathrm{X}}-\left(B_3-B_3^*\right)\right)\omega_{\mathrm{X}}t\sin\left(\omega_0t\right)\right.\right.\\\left.\left.+\left(B_3+B_3^*\right)\omega_{\mathrm{X}}^2t^2\cos\left(\omega_0t\right)\vphantom{\frac{2}{\omega_{\mathrm{X}}^3}}\right]\vphantom{\frac{1}{t^2}}\right\}.
\end{multline}
}
This expression, we are aware, is less than plain. Notice, however, that it features the now familiar ``seed'' of Fermi's golden rule, namely, the quantity
\begin{equation} \label{eq:GoldenSeed}
\frac{1}{t}\mathrm{i}\pi\,A_1\left(\mathrm{sgn}\left(t\right)+\frac{2}{\pi}\mathrm{Si}\left(\omega_0t\right)\right),
\end{equation}
found on the second line of (\ref{eq:IntheEnd}). Notice from (\ref{eq:CapitalAGalore}) that in the dipole limit $\omega_0/\omega_{\mathrm{X}}\rightarrow0$, one has $A_1=-\mathrm{i}\omega_0$, and one retrieves, from (\ref{eq:SomeChanceSurvival}), the decay constant given by Fermi's golden rule in the dipole approximation. Keeping $\omega_{\mathrm{X}}$ to its actual value, we find a relative error $\left|\mathrm{i}A_1-\omega_0\right|/\omega_0=\SI{1.33d-5}{}$. With $\omega_{\mathrm{X}}$ and thus $A_1$ kept to their actual value, our decay constant $\Gamma=\mathrm{i}\,2^{10}/\left(3^9\pi\right)c^2\alpha^3A_1$ matches that found by Facchi and Pascazio \cite{FacchiPhD,FacchiPascazio}, who treated the problem nonperturbatively, by finding the resolvent for the Hamiltonian (\ref{eq:Hamilton}) of the sytem in the Laplace domain and then getting back to the time domain. All the other terms in (\ref{eq:IntheEnd}) are short-time deviations from Fermi's golden rule.\\

\begin{figure}[t]
  \begin{center}
    \includegraphics[width=1\columnwidth]{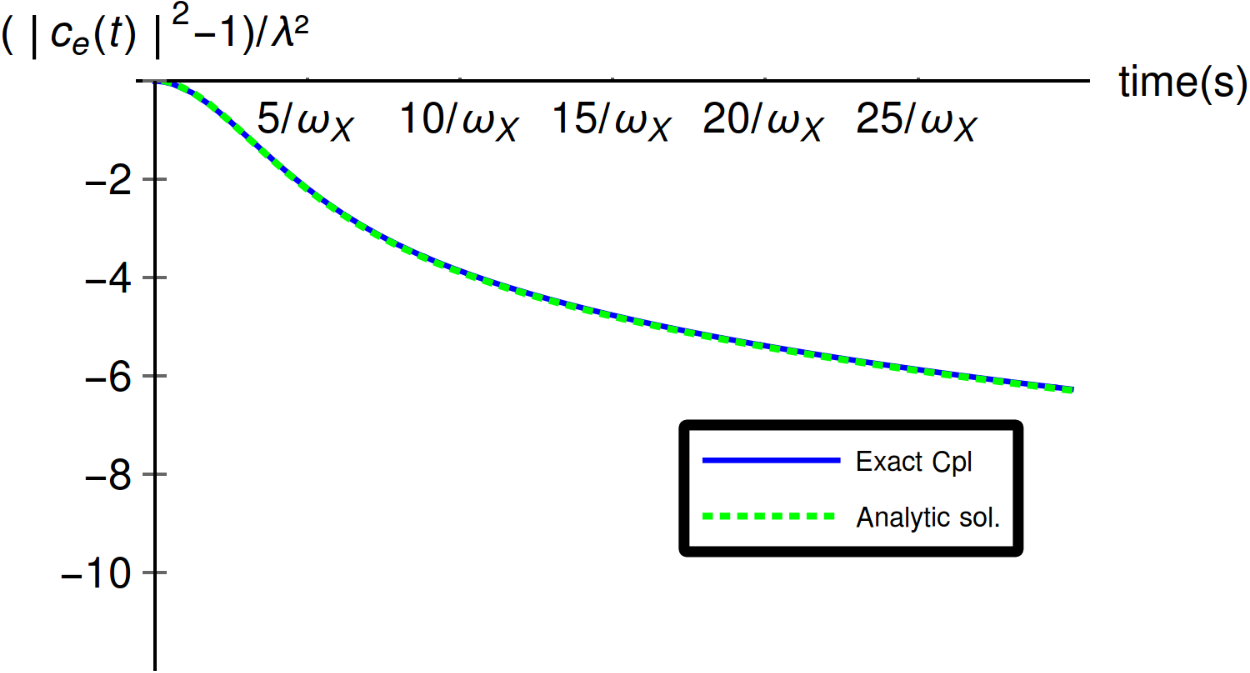}
  \end{center}
  \caption{Behaviour of the survival probability $\left|c_{\mathrm{e}}\left(t\right)\right|^2$ with the exact coupling as given by the perturbative solution (\ref{eq:IntheEnd}) (solid blue), and Facchi and Pascazio's exact solution \cite{FacchiPascazio} (dashed green). The dimensionless constant $\lambda^2=\frac{2}{\pi}\left(\frac{2}{3}\right)^9\alpha^3\simeq6.4\times10^{-9}$. Remember $1/\omega_{\mathrm{X}}=\SI{1.18d-19}{s}$. \label{fig:VeryVeryShort}}
\end{figure}

\begin{figure}[t]
  \begin{center}
    \includegraphics[width=1\columnwidth]{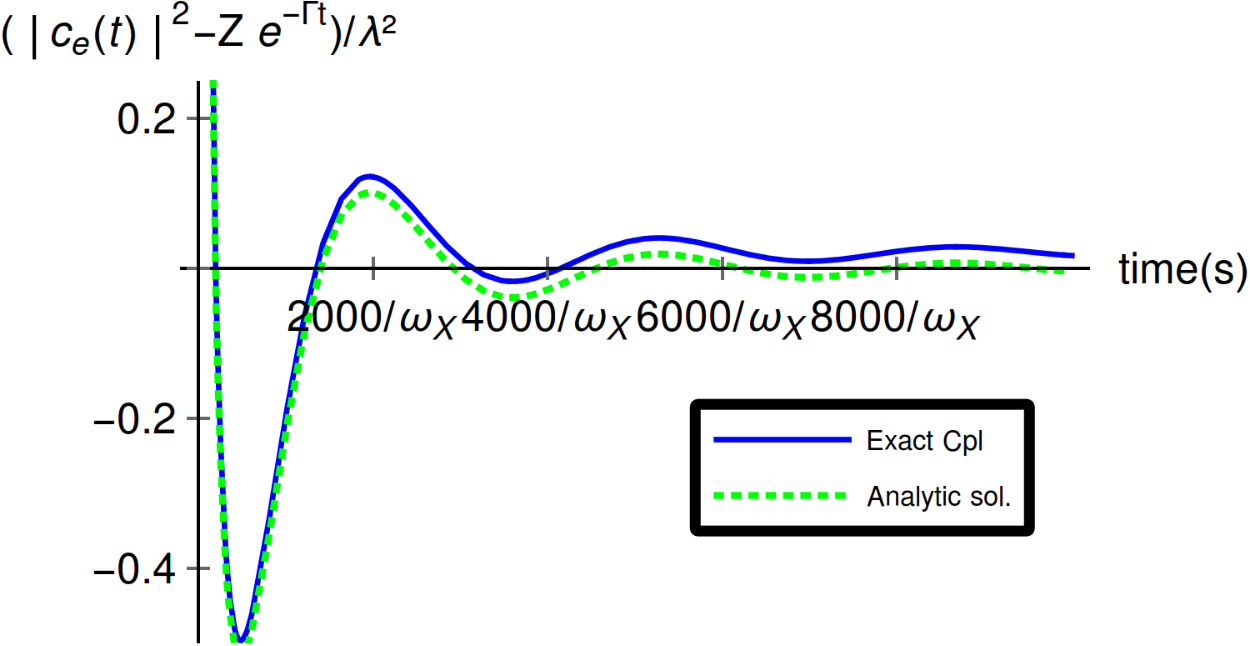}
  \end{center}
  \caption{Behaviour of the survival probability $\left|c_{\mathrm{e}}\left(t\right)\right|^2$ with the exact coupling as given by the perturbative solution (\ref{eq:IntheEnd}) (solid blue), and Facchi and Pascazio's exact solution \cite{FacchiPascazio} (dashed green). Here the exponential contribution $Z\mathrm{e}^{-\Gamma t}$ to the decay has been substracted, and only the nonexponential contribution thereto is plotted. The dimensionless constant $\lambda^2=\frac{2}{\pi}\left(\frac{2}{3}\right)^9\alpha^3\simeq6.4\times10^{-9}$. Remember $1/\omega_{\mathrm{X}}=\SI{1.18d-19}{s}$. Also note $Z\simeq1-4.39\,\lambda^2$ \cite{FacchiPascazio}. \label{fig:VeryShortCut}}
\end{figure}

It is not too hard, but very tedious to show from (\ref{eq:IntheEnd}), making use, of course, of (\ref{eq:CapitalAGalore}) as well as (\ref{eq:CapitalBGalore}), that the leading term in the Taylor series of $t^2\,I_F\left(t\right)$ at $t=0$ takes the very simple form $\left(\omega_{\mathrm{X}}t\right)^2/6$. Whence the survival probability at very short times:
\begin{equation} \label{eq:ZenoFacchi}
P_{\mathrm{surv}}\left(t\right)\underset{t\rightarrow0}{\sim}1-\frac{2^{10}}{3^{9}\pi}c^2\,\frac{\alpha^3}{6}\left(\omega_{\mathrm{X}}t\right)^2
\end{equation}
as deduced from (\ref{eq:SomeChanceSurvival}). This quadratic behaviour follows the lines of the usual Zeno regime \cite{MisraSudarshan}. The particular short-time expansion (\ref{eq:ZenoFacchi}) has also been obtained by Facchi and Pascazio \cite{FacchiPhD,FacchiPascazio}. As shown on Figs.~\ref{fig:VeryVeryShort} and \ref{fig:VeryShortCut}, the agreement between our perturbative treatment and the exact solution is very good. We can therefore use our method to investigate the short-time behaviour of this system in more detail.\\

\begin{figure}[t]
  \begin{center}
    \includegraphics[width=1\columnwidth]{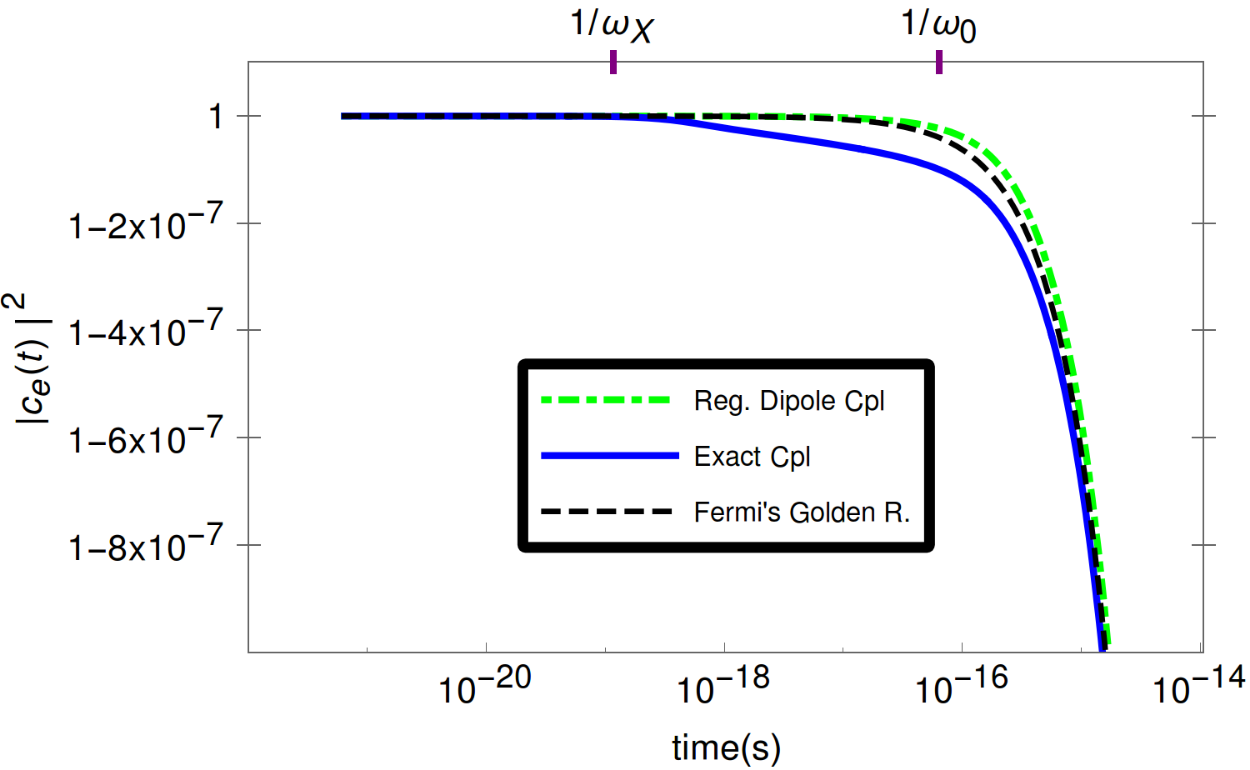}
  \end{center}
  \caption{Decay of the survival probability $\left|c_{\mathrm{e}}\left(t\right)\right|^2$ with the regularised dipole-approximated coupling (plotted for exhaustiveness) as given by expression (\ref{eq:RegOnly}) (dot-dashed green), the exact coupling as given by expression (\ref{eq:IntheEnd}) (solid blue) and Fermi's golden rule (dashed black). The time axis is logarithmic. \label{fig:AlmostFermi}}
\end{figure}

The survival probability (\ref{eq:SomeChanceSurvival}) is plotted in Fig.~\ref{fig:AlmostFermi}. We see that the deviation from Fermi's golden rule is of order $\SI{d-8}{}/\SI{d-7}{}$. For the sytem under study here, the golden rule is thus valid to an excellent approximation.\\

\begin{figure}[t]
  \begin{center}
    \includegraphics[width=1\columnwidth]{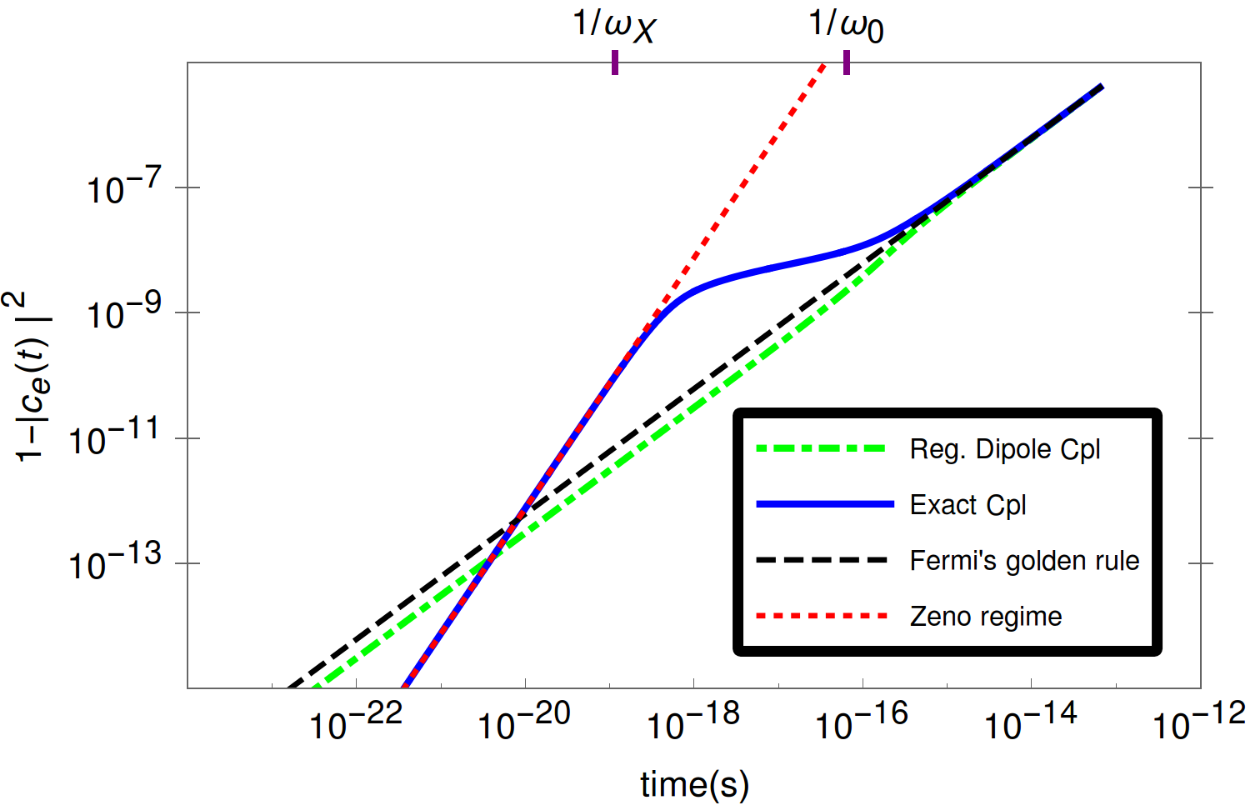}
  \end{center}
  \caption{Behaviour of the decay probability $1-\left|c_{\mathrm{e}}\left(t\right)\right|^2$ with the regularised dipole-approximated coupling (plotted for exhaustiveness) as given by expression (\ref{eq:RegOnly}) (dot-dashed green), the exact coupling as given by expression (\ref{eq:IntheEnd}) (solid blue), Fermi's golden rule (dashed black), and the Zeno behaviour (\ref{eq:ZenoFacchi}) (dotted red). Both axes are logatithmic. \label{fig:ZenoFermi}}
\end{figure}

To illustrate the transition between the Zeno regime (\ref{eq:ZenoFacchi}) and Fermi's golden rule as predicted by our treatment, we plot in Fig.~\ref{fig:ZenoFermi} the decay probability as given by (\ref{eq:IntheEnd}) as well as the short-time expansion (\ref{eq:ZenoFacchi}) and the linear prediction of the golden rule. Note that the transition between the Zeno and Fermi regimes takes place around $\SI{d-17}{s}$ after the start of the decay and that after $\SI{d-15}{s}$, the behaviour of the system is completely undistinguishable from that predicted by the golden rule.

\section{Conclusion} \label{sec:Ccl}

We have verified, in the framework of first order time-dependent perturbation theory, that Fermi's rule that predicts a linear decay of the survival probability can indeed, for the $2\mathrm{p}-1\mathrm{s}$ transition in atomic Hydrogen, be called ``golden''. The maximal deviation therefrom that we found is of order $\SI{d-8}{}/\SI{d-7}{}$, a clear-cut endorsement. It is a success for a ``rule'' which, as was argued in sect.~\ref{subsec:StartingHere}, is derived with the help of several, not obviously compatible conditions. As such, we think that much larger deviations from the golden rule could be found in other systems. As far as first-order perturbation theory goes, the crucial expression is the square cardinal sine integral in (\ref{eq:CardSineSurv}). It tells us that with an enhanced coupling between an atom or another effective two-level system and (electromagnetic) modes which are off-resonant with the transition frequency of the two-level system, one would witness more important deviations from the golden rule.\\

In \cite{FacchiPascazio} Facchi and Pascazio also raised the question of the experimental observability of Zeno deviations from the golden rule. They considered the ratio between the Zeno time $\tau_{\mathrm{Z}}$, where $P_{\mathrm{surv}}\left(t\right)\underset{t\rightarrow0}{\sim}1-\left(t/\tau_{\mathrm{Z}}\right)^2$, and the lifetime $1/\Gamma$ of the excited level as the relevant parameter for the observability. We argue that the relevant ratio is that between the ``cutoff time'' $\tau_{\mathrm{X}}$ and the Zeno time $\tau_{\mathrm{Z}}$. The cutoff time is understood to be defined so that after $\tau_{\mathrm{X}}$, the system exits the Zeno regime in which the survival probability decays quadratically. Therefore, at $t=\tau_{\mathrm{X}}$, we have $P_{\mathrm{surv}}\left(t\right)=1-\left(\tau_{\mathrm{X}}/\tau_{\mathrm{Z}}\right)^2$, and the strength of the Zeno effect is given by $\tau_{\mathrm{X}}/\tau_{\mathrm{Z}}$. This is confirmed by looking at Fig.~\ref{fig:ZenoFermi}, which shows that the maximal discrepancy between the predictions of the golden rule and the actual dynamics of the system is reached aproximately at the moment when the system exits the Zeno regime. There is a general method to obtain the Zeno time $\tau_{\mathrm{Z}}$, which is centred \cite{FacchiPhD} on the computation of the expectation value of the squared Hamiltonian $\hat{H}^2$ of the system in the initial state. On the other hand, it is difficult to obtain the cutoff time $\tau_{\mathrm{X}}$ without solving -at least perturbatively, as we did here- the dynamics of the system. Given the delicate nature of analytical approaches (see sect.~\ref{subsec:MultipoleCalc} as well as \cite{FacchiPhD,FacchiPascazio}) for this particular transition, which is much simpler than many other transitions to describe theoretically, we suspect that the best way to evaluate the cutoff time for a general transition would be by numerical evaluation of the integral on the right-hand side of (\ref{eq:CardSineSurv}). We showed in \cite{EdouardArXiv} that the ratio $\tau_{\mathrm{X}}/\tau_{\mathrm{Z}}$ scales, for hydrogen-like atoms with $Z$ protons, like $Z$, a favourable scaling for the observability of the Zeno regime.\\

\begin{figure}[t]
  \begin{center}
    \includegraphics[width=1\columnwidth]{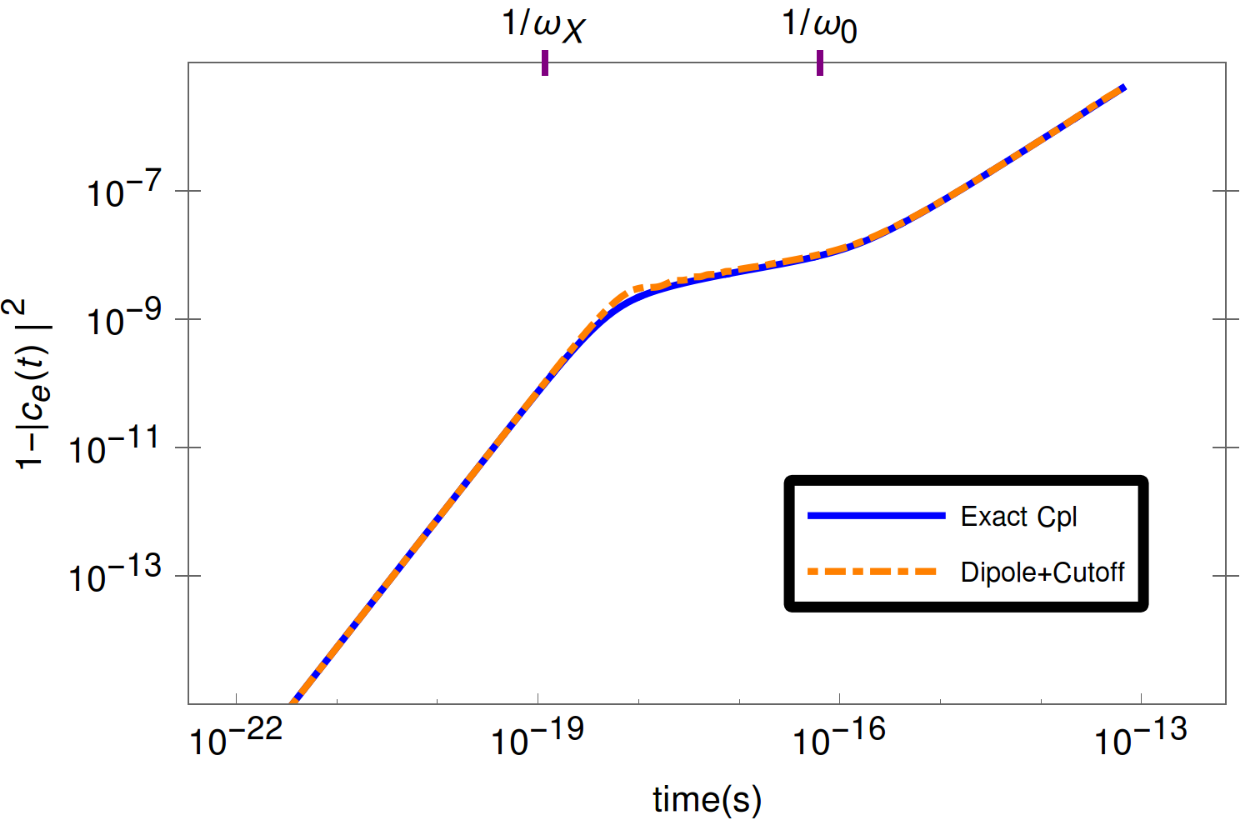}
  \end{center}
  \caption{Behaviour of the decay probability $1-\left|c_{\mathrm{e}}\left(t\right)\right|^2$ with the exact coupling as given by expression (\ref{eq:IntheEnd}) (solid blue) and the dipole-approximated coupling with cutoff frequency $\omega_{\mathrm{C}}=\omega_{\mathrm{X}}/\sqrt{3}$ as given by expression (\ref{eq:Trunk}) (dot-dashed orange). Both axes are logatithmic. \label{fig:CheatandWin}}
\end{figure}

Our investigations also shed light on the dipole approximation. We have seen that while the regularisation procedure of sect.~\ref{subsec:DipoleCalc} provides a nicely cutoff-independent treatment of the problem in the framework of the dipole approximation, and yields a result which is in agreement with Fermi's golden rule at ``long times'', the predictions it yields on the very short time dynamics of the system are inadequate. Namely, it does not provide the correct dynamics in the Zeno regime, as seen on Fig.~\ref{fig:ZenoFermi}. We might ask, however, how the predictions of the dipole approximation fare when the regularisation is performed more directly -and, arguably, less elegantly- via the introduction of a cutoff, as presented in sect.~\ref{subsec:DipoleDsc}. As expected from the usual Zeno dynamics \cite{MisraSudarshan}, the very short time behaviour yielded by the truncated integral (\ref{eq:Trunk}) is a quadratic decay in time. Namely
\begin{equation} \label{eq:TrunkEarly}
1-\frac{2^{10}}{3^{9}\pi}c^2\,\alpha^3t^2\hspace{-2.5pt}\int_0^{\omega_{\mathrm{C}}}\hspace{-7.5pt}\mathrm{d}\omega\,\omega\sin\!\mathrm{c}^2\left[\left(\omega_0-\omega\right)\frac{t}{2}\right]\underset{t\rightarrow0}{\sim}1-\frac{2^{10}}{3^{9}\pi}c^2\frac{\alpha^3}{2}\left(\omega_{\mathrm{C}}t\right)^2.
\end{equation}
This can either be seen by computing the Taylor series of the right-hand side of (\ref{eq:Trunk}), or, much more directly, by taking $t=0$ in the integral on the left-hand side of (\ref{eq:TrunkEarly}). Now, remember that in the case of the exact coupling, the Zeno behaviour is given by (\ref{eq:ZenoFacchi}). One can then choose the cutoff frequency of the dipole approximation so that the very short time predictions of the dipole approximation, with cutoff, match the exact short time dynamics of the system. Comparison of (\ref{eq:ZenoFacchi}) with (\ref{eq:TrunkEarly}) shows that a perfect match is reached if we choose
\begin{equation} \label{eq:UMadBro}
\omega_{\mathrm{C}}\equiv\frac{\omega_{\mathrm{X}}}{\sqrt{3}}=\frac{\sqrt{3}}{2}\frac{c}{a_0}\simeq .866\frac{c}{a_0}.
\end{equation}
In Fig.~\ref{fig:CheatandWin} we compare the predictions of the dipole approximation, with the carefully picked cutoff frequency (\ref{eq:UMadBro}), with the predictions obtained with the exact atom-field coupling. It is interesting, and quite impressive, that while the dipole approximation here was made to fit, by a simple choice of the cutoff frequency, the exact Zeno dynamics of the system, we see that with our choice for the cutoff, we obtain an excellent agreement during the transition between the Zeno and Fermi regimes\footnote{We also obtain an excellent agreement in the Fermi regime, but that was to be expected. The agreement is not perfect, though, as the decay constant in the exact and dipole coupling are slightly different. See the discussion below (\ref{eq:GoldenSeed}).}. Hence we can conclude that for the $2\mathrm{p}-1\mathrm{s}$ transition in atomic Hydrogen, the dynamics of the system is very well described at all times within the framework of the dipole approximation, if one makes the ``correct'' choice (\ref{eq:UMadBro}) for the cutoff.

\section*{Acknowledgments}
Vincent Debierre acknowledges support from CNRS (INSIS doctoral grant). Thomas Durt acknowledges support from the COST 1006 and COST 1043 actions. We thank Pr. Édouard Brainis for helpful discussions and valuable suggestions on the presentation of our results.

\bibliography{SlowDecay}

\end{document}